\documentclass[unnumsec,webpdf,modern,medium]{oup-authoring-template}

\usepackage{comment}
\usepackage{lipsum} 
\graphicspath{{Fig/}}

\theoremstyle{thmstyleone}%
\theoremstyle{thmstyletwo}%
\theoremstyle{thmstylethree}%

\begin{document}

\journaltitle{}
\copyrightyear{}
\appnotes{Review article}

\firstpage{1}


\title[Non-bilaterians as Model Systems for Tissue Mechanics]{Non-bilaterians as Model Systems for Tissue Mechanics}

\author[1,$\ast$]{Setareh Gooshvar\ORCID{0000-0001-5571-8456}}
\author[1,$\ast$]{Gopika Madhu\ORCID{0009-0007-9581-8022}}
\author[1]{Melissa Ruszczyk\ORCID{0000-0003-2430-1847}}
\author[1,2,3,$\ast \ast$]{Vivek N. Prakash\ORCID{0000-0003-4569-6462}}

\authormark{Gooshvar, S. et al.}

\address[1]{\orgdiv{Department of Physics, College of Arts and Sciences}, \orgname{University of Miami}, \state{FL}, \country{U.S.A.}}
\address[2]{\orgdiv{Department of Biology, College of Arts and Sciences}, \orgname{University of Miami}, \state{FL}, \country{U.S.A.}}
\address[3]{\orgdiv{Department of Marine Biology and Ecology, Rosenstiel School of Marine, Atmospheric and Earth Science}, \orgname{University of Miami},  \state{FL}, \country{U.S.A.}}

\corresp[$\ast$]{Equal contribution~~}

\corresp[$\ast \ast$]{Corresponding author: \href{vprakash@miami.edu}{vprakash@miami.edu}}




\abstract{In animals, epithelial tissues are barriers against the external environment, providing protection against biological, chemical, and physical damage. Depending on the animal's physiology and behavior, these tissues encounter different types of mechanical forces and need to provide a suitable adaptive response to ensure success. Therefore, understanding tissue mechanics in different contexts is an important research area. Here, we review recent tissue mechanics discoveries in a few early-divergent non-bilaterian animals -- \textit{Trichoplax adhaerens}, \textit{Hydra vulgaris}, and \textit{Aurelia aurita}. We highlight each animal's simple body plan and biology, and unique, rapid tissue remodeling phenomena that play a crucial role in its physiology. We also discuss the emergent large-scale mechanics that arise from small-scale phenomena. Finally, we emphasize the enormous potential of these non-bilaterian animals to be model systems for further investigation in tissue mechanics.}  


\maketitle

\section{Introduction}

\begin{figure*}[h]
        \centering\includegraphics[width=14 cm]{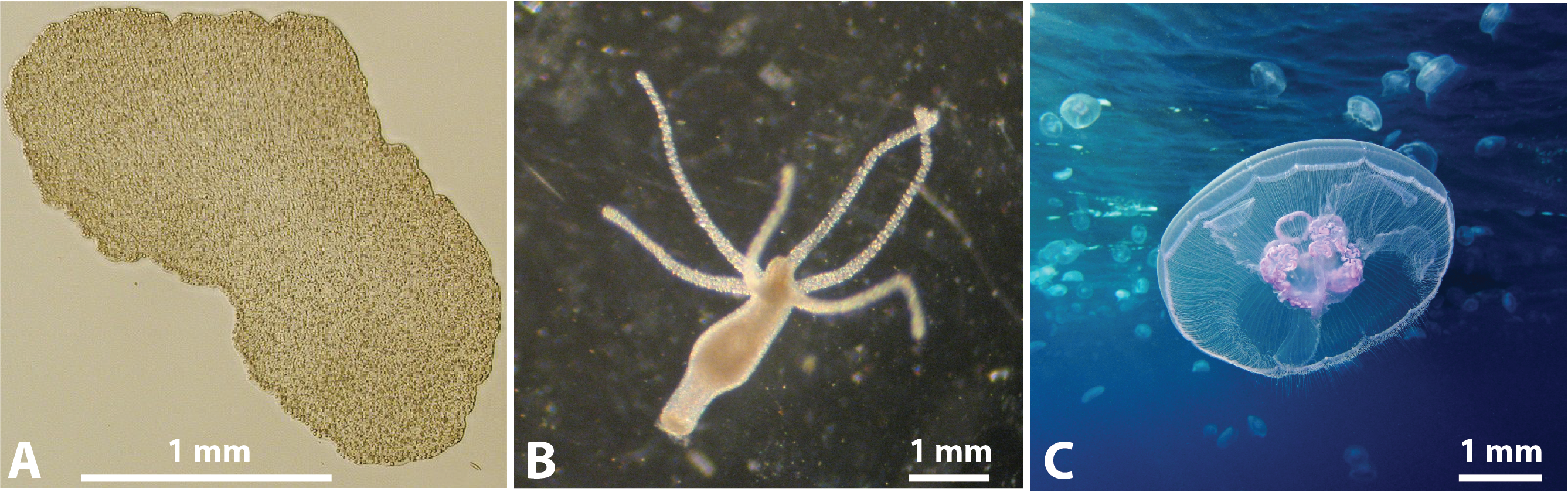}
        \caption{Non-bilaterian animals are excellent model systems to study tissue mechanics. (A) \textit{Trichoplax adhaerens} of phylum Placozoa, Image courtesy of  Oliver Voigt / Wikimedia Commons / CC-BY-SA-3.0. (B) \textit{Hydra vulgaris} of phylum Cnidaria, Image courtesy of Stephanie Guertin / Wikimedia Commons / CC-BY-SA-3.0. (C) \textit{Aurelia aurita} (Moon jellyfish) of phylum Cnidaria, Image courtesy of Alexander Vasenin / Wikimedia Commons / CC-BY-SA-3.0.}
        \label{intro_figure}
\end{figure*}

Epithelial tissues in animals are subjected to different types of mechanical forces during their entire life cycle. In order to foster the animal's survival and success, these tissues must be able to respond, adapt and withstand external forces. Thus, an important goal of tissue mechanics research is to apply the principles of mechanics to characterize and quantify the mechanical properties and understand the response of biological tissues~\cite{Fung1990}. A key question is to understand how the properties of individual cells and their collective interactions give rise to emergent mechanical properties of the tissue.

Tissue mechanics also plays an important role in biological processes such as development and physiology. Significant advances have already been made in tissue mechanics research focused on human biomedical applications~\cite{Fung1990,Park2015}; as well as in the context of developmental processes in model animals such as the fruit fly, \textit{D. melanogaster}, and zebrafish, \textit{D. rerio}~\cite{lecuit2013,lecuit2007cell,blankenship2006multicellular,he2014,Mongera2018}. In-vitro cell-culture systems have also been a popular model for tissue mechanics studies~\cite{Charras2012,Latorre2018,Xi2019,Charras2022}. However, very little is known regarding the tissue mechanics of early divergent, non-bilaterian animals. Here, we highlight recent surprising tissue mechanics discoveries in non-bilaterian animals and their potential in opening up new questions and research directions~\cite{prakash2021motility,Carter2016,pnas.1502497112}. 

Non-bilaterian animals include some of the first multi-cellular animals that evolved from unicellular organisms. These are the most ancestral, earliest diverging animals that do not have a bilateral symmetry axis. The non-bilaterian phyla of animals includes Porifera, Placozoa, Ctenophora, and Cnidaria~\cite{Dunn2014}. These non-bilaterians have been the focus of several recent biological studies, but have received very little attention from a biophysical perspective. The objective of this review is to attract the attention of biomechanicians and biophysicists and motivate them to work on these important animal systems. 

Here, we will focus on three non-bilaterians: the \textit{Trichoplax adhaerens} of Phylum Placozoa, \textit{Hydra vulgaris}, and \textit{Aurelia aurita} of phylum Cnidaria (Figure~\ref{intro_figure}). These non-bilaterian animals provide several key advantages to serve as model research animals for tissue mechanics -- they have all been successfully maintained in the lab, are soft and experimentally tractable, have simple body plans, and their tissues are suitable for live imaging. In these animals, the organs and organ systems are not as well developed and specialized as in bilaterians, so organization on the tissue level plays a prominent role  in their physiology and behavior. Here, our focus will be on fast-time scale (minutes to hours) tissue remodeling events and their mechanical processes. We will focus on tissue remodeling phenomena such as local cellular rearrangements, tissue fractures and their healing, in the context of physiological processes such as feeding, locomotion, reproduction, and repair. Important aspects such as growth and regeneration that take place over longer time-scales (several days/weeks) are neglected.

This review is organized as follows: In every system, we introduce the animal's biology and body plan organization. Then, we describe its tissue layers and their role in the animal's physiology before highlighting recently reported tissue mechanics phenomena in the animal. In the discussion section, we summarize and emphasize how tissue mechanics plays an important role in form-function relationships in these three animal systems. We also reveal how small-scale phenomena give rise to large-scale behaviors. Finally, we conclude by highlighting how non-bilaterians are excellent model systems to investigate the biophysics of tissues.

\section{Trichoplax}\label{sec1}
\begin{figure*}[h]
        \centering\includegraphics[width=15 cm]{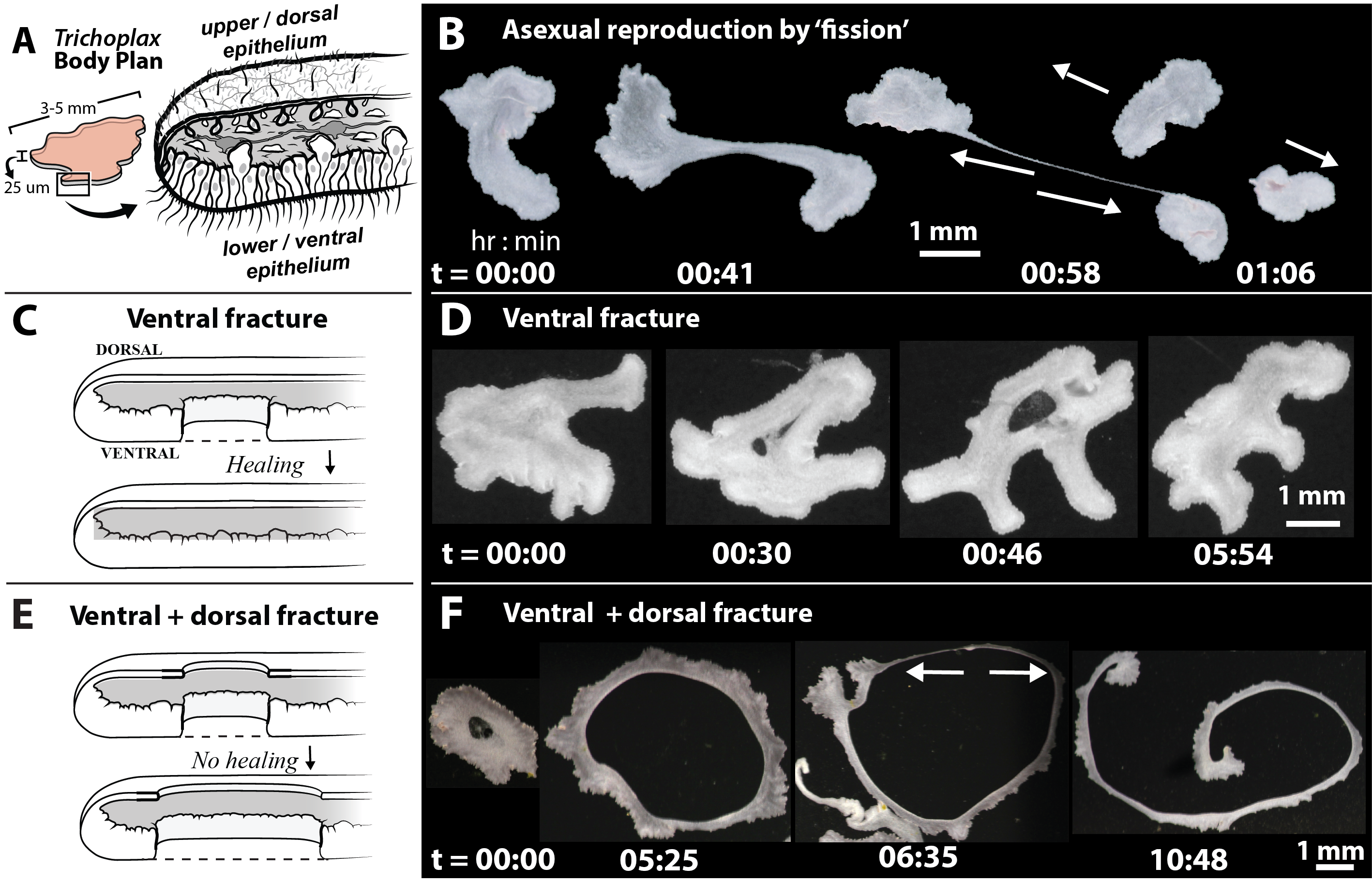}
        \caption{\textbf{Tissue mechanics in \textit{Trichoplax adhaerens}~\cite{prakash2021motility}:} \textbf{A.} Cartoon shows the flat body plan, and cross-section shows the upper/dorsal and lower/ventral epithelial tissue layers. \textbf{B.} Time-lapse image sequence of asexual/vegetative reproduction process; in the central region, tissues deform rapidly to become a narrow thread that eventually breaks. \textbf{C, D.} Schematic and time-lapse image sequence reveals formation of a fracture hole in the ventral epithelium, and its subsequent healing. \textbf{E, F.} Schematic and time-lapse image sequence show formation of a fracture hole in both ventral and dorsal epithelium. This type of fracture hole does not heal, but instead enlarges to form a thin edge that eventually breaks and gives rise to a string-like animal. Images in this figure are adapted and reproduced, from~\cite{prakash2021motility}.}
        \label{fig:Trichoplax}
\end{figure*}

The early-divergent marine animal \textit{Trichoplax adhaerens} (Figure~\ref{intro_figure}A) is the most well-studied member  of the basal animal phylum of Placozoa~\cite{Schierwater2018,Armon2018a,Smith2014a,Srivastava2008}. \textit{Trichoplax adhaerens} is one of the three species of animals in the entire phylum; and the two other animals were only found recently~\cite{eitel2018comparative,osigus2019polyplacotoma}. \textit{Trichoplax adhaerens} has been found in many parts of tropical oceans around the world \cite{Pearse2007FieldInteractions,eitel2013global}. \textit{Trichoplax adhaerens} is considered to be the simplest free-living animal since it consists of less than ten cell types~\cite{Smith2014a,romanova2021hidden}, and lacks neurons, muscles, extracellular matrix, and a basement membrane~\cite{Schierwater2018,Srivastava2008}. This animal has a simple and flat body plan, with a thickness of only 25 $\mu$m, but the animals can exhibit a large variation in width, ranging from $\sim$50 $\mu$m to 10 mm (Figure~\ref{intro_figure}A). 

\subsection{The Role of Tissue Layers in \textit{Trichoplax} Physiology}\label{subsec1}

In \textit{T. adhaerens}, the flat body plan consists of three tissue layers, an upper dorsal epithelium, a central layer of fiber cells, and a lower ventral epithelium. The dorsal epithelial cells have a thin and flat architecture, while the ventral epithelial cells have a columnar structure, and the two tissue layers are coupled at the edge of the animal. The epithelial cells are connected together only by adherens junctions, and no tight junctions have been found~\cite{Smith2016}. Although both tissue layers have monociliated cells, the cilia of ventral epithelial cells are unique since they can adhere to the bottom substrate, providing the organism sufficient traction forces to be able to walk and generate push/pull forces~\cite{bull2021a}. \textit{Trichoplax adhaerens} does not have a fixed shape like other animals. Instead, the animal is constantly changing shape in an amorphous manner driven by ciliary traction with the bottom substrate. The animals exhibit an extreme range of shape morphologies ranging from circular-like shapes to long elongated threads~\cite{prakash2021motility}. \textit{Trichoplax adhaerens} also reproduces by vegetative or asexual fission, resulting in two or more daughter animals~\cite{prakash2021motility,Eitel2011,Srivastava2008} (Figure~\ref{fig:Trichoplax}B). In these animals, epithelial tissue remodeling processes can play an important role in determining the organismal shape change dynamics and asexual reproduction.

\subsection{Tissue Remodeling via Cellular Rearrangements}\label{subsec2}

The vegetative (or asexual) reproduction process in \textit{Trichoplax adhaerens} begins when an individual animal forms two coherent regions that start pulling away from each other (Figure~\ref{fig:Trichoplax}B). The tissues in between the two pulling regions are subjected to mechanical forces (tension), and respond by undergoing a rapid thinning deformation to form a thin narrow thread (taking less than 1 hour). From a materials science viewpoint, this rapid thinning process resembles a `ductile' material transformation process, where `ductility' refers to the ability of a material to be drawn into thin wires. This rapid thinning process involves local tissue remodeling via fast time-scale cellular rearrangement mechanisms~\cite{prakash2021motility}. If the two opposing parts of the animal are able to generate sufficient traction forces, eventually the thin thread will break at the length-scale of a single cell and result in the formation of two or more daughter animals, thereby completing the reproduction process~\cite{prakash2021motility,Eitel2011,Srivastava2008}. The smaller daughter animals will grow in size and again undergo asexual fission when their size reaches $>$2-3mm ~\cite{prakash2021motility}.

\subsection{Tissue Remodeling via Fractures}\label{subsec3}

\textit{Trichoplax adhaerens} can grow to large sizes ($>$2-3mm), and these larger animals are capable of executing extreme shape changes, making them a very interesting model animal for tissue mechanics. Larger animals can generate larger traction forces due to their motility, and this leads to very surprising shape morphologies, such as fracture holes and their healing dynamics, and long string-like animals ~\cite{prakash2021motility} (Figure~\ref{fig:Trichoplax}C-F). These fracture holes can be rapidly induced in the bulk of their ventral tissues solely due to motility-induced tensile or shear mechanical forces at the organismal scale~\cite{prakash2021motility} (Figure~\ref{fig:Trichoplax}C). These fracture holes begin at small scales as micro-fractures and rapidly coalesce (in about half an hour) to form large stable ventral holes that are visible at the organismal scale (Figure~\ref{fig:Trichoplax}D). From a materials science viewpoint, this fracture formation and growth process resembles a `brittle' material transformation process, where `brittleness' refers to the tendency of a material to break. In many cases, it was observed these ventral holes can also rapidly heal themselves (in about half an hour) if the hole edges came into contact due to the animal's motility (Figure~\ref{fig:Trichoplax}D). 

Sometimes, the ventral holes do not heal and the dorsal epithelium also sustains a fracture hole right above the ventral hole (Figure~\ref{fig:Trichoplax}E,F). The animal is now left with a through-hole inside it, and the two tissue layers seal themselves to form an permanent edge inside the animal (preventing any further healing). This animal will now have the geometry of a toroid (like a donut). Over a time-scale of $\sim$10 hours, the inside hole diameter keeps increasing until one of the edges becomes thin and eventually breaks, giving rise to long string-like animals (Figure~\ref{fig:Trichoplax}F). From an organismal morphology perspective, fractures enable the fastest topological transformations from a circular shape to a long string-like shape, compared to a shape change mechanism that relies on cellular rearrangements.  

These tissue remodeling processes in \textit{T. adhaerens} were further investigated using in-silico tissue models~\cite{prakash2021motility}. The simplified heuristic two-dimensional (sheet) model of ventral epithelium consisted of soft balls (representing cells) that were connected to each other by springs (representing adhesion bonds). The tissue model was subjected to tensile loading (pulling forces) and the resulting tissue response was studied. We emphasize here that the focus is on tissue remodeling processes at fast time-scales (minutes) and the long term effects of growth (several hours) has been neglected. It was found that the two key parameters governing tissue response were the pulling force and the length at which the springs break. Simulations exploring a wide range of these two parameters resulted in a phase diagram that revealed an elastic--ductile--brittle transitions in the material properties. The model captured experimental observations faithfully. Thus confirming that elastic--ductile tissue transitions occur during the local cellular rearrangement process when the animals are dividing by vegetative fission (Figure~\ref{fig:Trichoplax}B); and that elastic--brittle transitions indeed occur when the animals sustain tissue fractures (Figure~\ref{fig:Trichoplax}D,F) during organismal shape change. Hence, \textit{T. adhaerens} is an excellent model system for further investigations in tissue mechanics, since solely mechanical forces can give rise to the tissue remodeling processes that play a critical role in their life-cycle~\cite{prakash2021motility}. 

\section{Hydra}\label{sec2}

\begin{figure*}[h]
        \centering\includegraphics[width=11 cm]{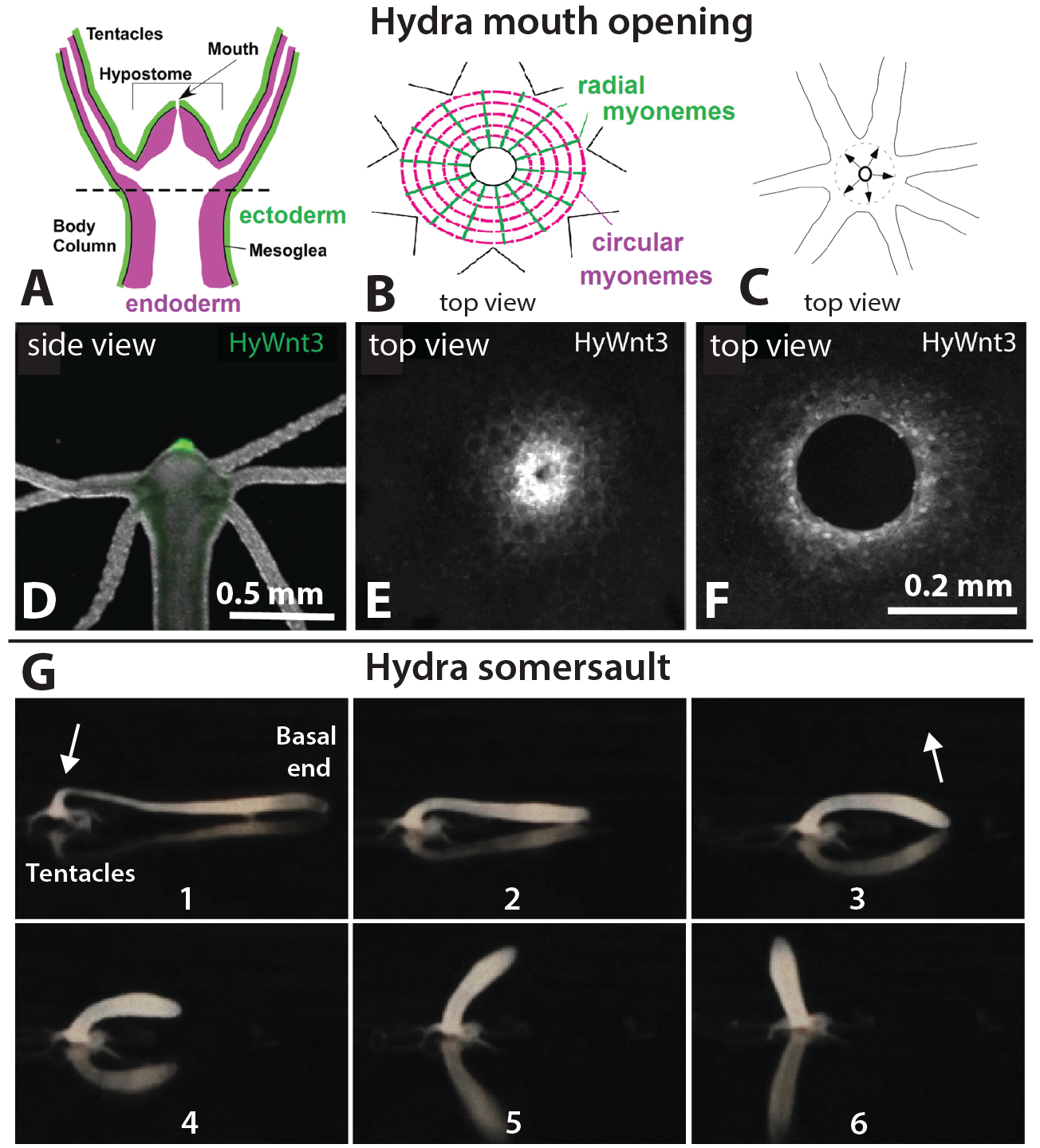}
        \caption{\textbf{Tissue mechanics in \textit{Hydra vulgaris}:} \textbf{A, D} Cartoon and brightfield/fluorescence image shows the side view of \textit{Hydra}; the body plan consists of two tissue layers: a thinner ectoderm (green outline) and a thicker endoderm (purple outline). The mouth lies in the center of a ring of tentacles and together they compose the \textit{Hydra}'s ``shoulder", i.e. the region above the dotted line \cite{Carter2016}. \textbf{B, E.} When closed, \textit{Hydra}'s mouth is composed of radial and circular contractile units called myonemes \cite{Carter2016}. \textbf{C, F.} When mouth opening occurs, the contractile myonemes assist in providing the opening force necessary to tear the ectodermal and endodermal tissue layers, before sealing them once again \cite{Carter2016}. \textbf{G.} The \textit{Hydra} somersault driven by differential tissue stiffness, is composed of three general stages. In Stage I (panels 1-2) the \textit{Hydra} body column extends while tentacles anchor to the substrate. In Stage II (panels 3-4), it disconnects the basal end from substrate and raises the body column. In Stage III (panels 5-6), the body column is completely raised \cite{Naik2020,Wang2023}.
        Images in this figure are adapted and reproduced, with permission, from~\cite{Carter2016, Naik2020}.} 
        \label{fig:hydra}
\end{figure*}

\textit{Hydra vulgaris} is a freshwater polyp of the Cnidaria phylum which exhibits the characteristic radial symmetry, cnidocytes, and body plan derived from two germ layers \cite{McLaughlin2017}. \textit{Hydra}'s body plan consists of a two-layered tube body and a mouth composed of a ring of tentacles and dome-shaped hypostome (Figure~\ref{fig:hydra} A,D). 
Although \textit{Hydra} is an invertebrate composed of two epithelial layers, the endoderm and ectoderm, separated by an extracellular mesoglea. Cells in the tissues of the body column cycle continuously with those of the head and foot regions, maintaining the equilibrium between cell production and loss~\cite{Wang2023, Galliot2006}. 

\subsection{The Role of Tissue Layers in \textit{Hydra} Physiology}\label{subsec4}

In addition to the dynamic nature of its tissues, the existence of a differential thickness between the thinner ectodermal layer and the thicker endodermal layer is of particular importance. This difference in thickness is most pertinent when examining the \textit{Hydra} mouth. When it is closed, the mouth is a continuous epithelial sheet sealed with septate junctions, which were first seen in the model organism \textit{Drosophila}, and appear as a ladder-like junction between two cells \cite{Banerjee2006,Izumi2014}. These septate junctions thus act as intercellular connectors functioning to merge adjacent epidermal cells with inner, luminal edges of cells \cite{Carter2016, Hand1972}. 
 
\subsection{Mouth Opening Dynamics}\label{subsec5}

When the \textit{Hydra} opens its mouth, it must tear a hole through the epithelial tissues at each instance of opening. It does so exclusively through the viscoelastic deformation of cells. This was confirmed via cell shape analysis and by tracking individual cells during mouth openings. It was observed that cells were not undergoing rearrangement and were instead conserving existing cellular contacts \cite{Carter2016}. The \textit{Hydra} must overcome the mouth opening force that exists while the endoderm and ectoderm are sealed. When the mouth is closed, the septate junctions act to connect the cells in both epithelial sheets. In the ectoderm, the sealing force to be overcome is solely that of the septate junctions while the endoderm needs to additionally include myonemes, which are circularly oriented contractile structures (Figure~\ref{fig:hydra} B,E). Both of these closing forces must be exceeded by the opening force to lead to successful mouth opening. These sealing forces have been estimated to be on the order of several nN \cite{Carter2016}. 

Although the force required must be estimated, the kinematics of the mouth opening can be fully characterized by a logistic equation \cite{Carter2016}. Comparing the normalized mouth opening area to the time axis resembles a logistic curve which can then be time-shifted and normalized to the ectoderm and endoderm mouth opening area. This results in a modified logistic equation which accounts for the normalized area of the mouth as a function of time, and various fit parameters to allow for the full capture of the kinematics of mouth opening for both ectoderm and endoderm separately (Figure~\ref{fig:hydra} C,F) \cite{Carter2016}. 

\subsection{The Somersaulting \textit{Hydra}}\label{subsec6}

Tissue mechanics also plays an important role for \textit{Hydra}'s locomotion. A differential stiffness within its body enables it to perform the “\textit{Hydra} somersault” \cite{Mackie1974, Han2018-om}. The \textit{Hydra} somersault occurs in three general stages: in stage 1, the body column is stretched, and the tentacles hold on to the substrate (Figure~\ref{fig:hydra} G:1,2). In stage 2, the basal end is released (Figure~\ref{fig:hydra} G:3,4). In stage 3, the body column contracts and is lifted, transporting the base to a new location in the direction of the \textit{Hydra}'s motion (Figure~\ref{fig:hydra} G:5,6). 

This type of movement is only possible due to the difference in the local mechanical properties of the \textit{Hydra}’s body column, specifically that of a 3:1 ratio in the Young’s modulus between the shoulder region and the body column \cite{Naik2020}. Thus, when the \textit{Hydra} moves from one place to another, it utilizes the differential stiffness of the body column to enable efficient transfer of mechanical energy stored in stretching to bending \cite{Naik2020, Wang2023}. 

Hence, \textit{Hydra} is an interesting animal that employs tissue  fracture mechanics for its mouth opening and differential stiffness activity for its locomotion.

\section{Jellyfish}\label{sec3}

\begin{figure*}[h]
        \centering\includegraphics[width=15 cm]{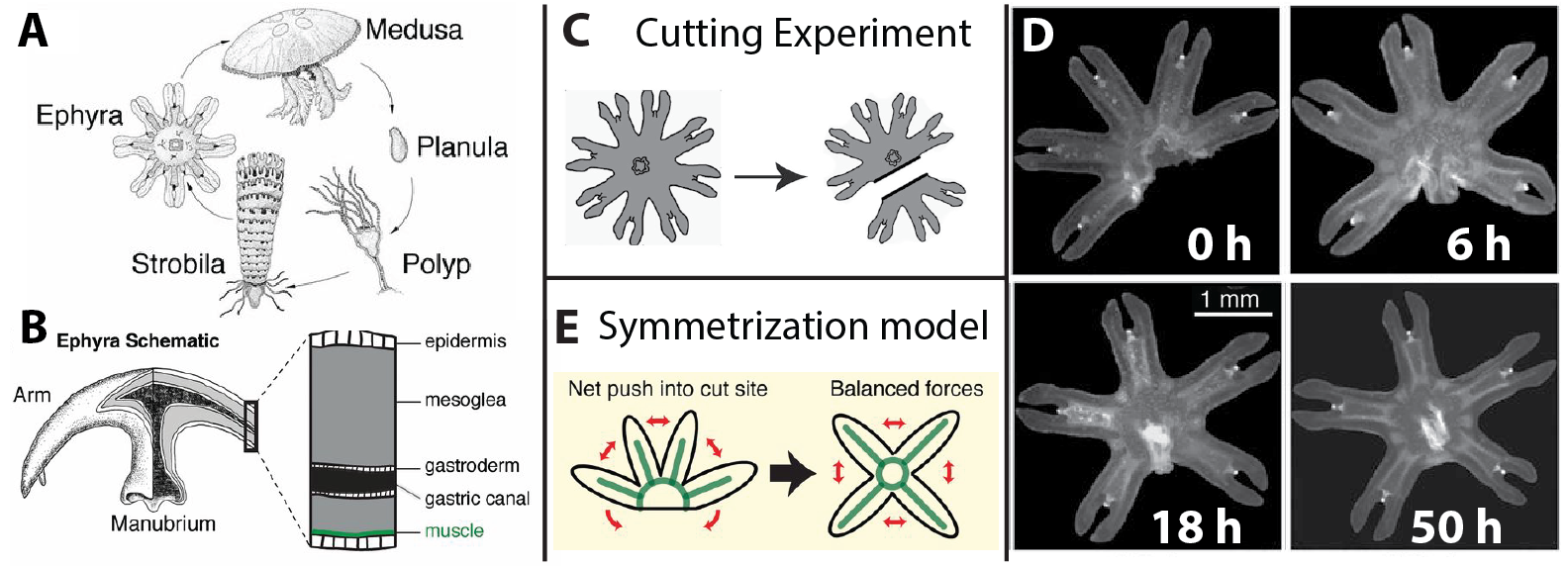}
        \caption{\textbf{Tissue mechanics in the Moon Jellyfish \textit{Aurelia artura}~\cite{pnas.1502497112}:}  \textbf{A.} Cartoon shows the Life-cycle: Fertilized eggs develop into planulae (larval stage) which then develop into a non-motile polyp form. Next, the polyps transform into strobilae and release free-swimming ephyrae (juvenile stage), which later develop into medusae (adult stage). \textbf{B.} Cross-section of ephyra: The epithelium consists of an epidermis, gastric layer and mesoglea -- a viscoelastic substance that fills the volume between the epidermis and gastric lining. \textbf{C.} Schematic showing the experimental amputation of arms in ephyra. \textbf{D.} Time series images of shape symmetrization (without regeneration) of the five-armed amputated ephyra. \textbf{E.} Schematic of the symmetrization model where arms relocate to a new position until the forces are rebalanced, i.e. until symmetry is regained. Images in this figure are adapted and reproduced, from~\cite{pnas.1502497112} / Creative Commons / CC BY-NC-ND.}
        \label{fig:Jellyfish}
\end{figure*}

While the previous section discusses tissue fracture mechanics of the Cnidarian \textit{Hydra} polyp, this section discusses the Cnidarian Medusozoans, commonly known as jellyfish. The Scyphozoan jellyfish have a life cycle with two adult forms, with one being the sexually reproducing fully motile medusae and the other form being the asexually reproducing sessile polyp (Figure \ref{fig:Jellyfish} A) \cite{10.1007/978-94-010-0722-1_19}.

Jellyfish are morphologically complex invertebrates with striated muscles. Despite lacking a functional brain, they possess a complex radially distributed neural and sensory systems that help them detect light and odour, enabling quick responses to stimuli \cite{SATTERLIE2011}. The cross-sectional view of ephyra reveals three layers: the epidermis, mesoglea, and gastrodermis. The epidermis, which is the outer layer, contains the neural net. The endodermis-derived inner layer, the gastrodermis, lines the gastric cavity. The region between the ectoderm and the endoderm is filled by mesoglea, a viscoelastic substance (Figure \ref{fig:Jellyfish} B) \cite{ABRAMS20161}.

The habitat of jellyfish is versatile, ranging across varying temperatures and depths. 
Though jellyfish can be considered top predators in their ecosystems, many predators, including other jellyfish, attack jellyfish opportunistically \cite{predators_jellyfish}. Therefore injury via amputation is common among jellyfish, and recovery from such damage could either be from re-organization or regeneration. Jellyfish have been widely studied for their regenerative abilities and symmetrization, specifically the medusae and ephyrae (juvenile jellyfish) stages of their life cycle. The focus here will be on the fast timescale process of tissue symmetrization.

\subsection{The Role of Tissue Layers in \textit{Aurelia} Physiology}\label{subsec7}

A way by which jellyfish can repair sustained damage that is not regenerative is through the restoration of symmetry after tentacle amputation, which occurs purely through responses to mechanical stress. The moon jellyfish (\textit{Aurelia aurita}) is an excellent example of symmetry restoration after the removal of one of its arms, known as tenticulocytes  \cite{ABRAMS20161}. An undamaged moon jellyfish has eight tenticulocytes forming its swimming apparatus as the arms symmetrically pulsate. The striated musculature of the arms connected to the jellyfish's pulsating "bell" through epithelial tissue drives the recovery and power strokes of the pulsations as the jellyfish swims. The mesoglea's elastic properties are used in the recovery stroke to restore the bell to its original shape \cite{JELLYFISHPROPUSION}. Because the architecture of the musculature is so symmetrically interlinked with the umbrella, any loss of symmetry, such as the removal of arms, creates an imbalance in forces during pulsations. Since radial symmetry is essential for propulsion, when any number of these arms are removed in a moon jellyfish ephyra (juvenile jellyfish), a recovery of radial symmetry is expected and observed (Figure \ref{fig:Jellyfish}) in a process termed as "symmetrization" by Abrams et al. \cite{ABRAMS20161}.\\

\subsection{\textit{\textit{Aurelia aurita}}-Symmetrization Model}

In the symmetrization model, Abrams et al.\cite{ABRAMS20161} suggest two repair strategies without increased cell proliferation in Cnidaria, one to restore lost parts and another to restore functional symmetry without restoring lost parts. The latter involves the process of symmetrization, where the arms eventually get rearranged to regain symmetry \cite{ABRAMS20161}.

The experiment involving amputation of arms from individual \textit{Aurelia} ephyrae (as shown in Figure \ref{fig:Jellyfish} C) resulted in wound closure of the injured site in three hours.  Within 18 hours, full symmetrization with the manubrium relocating to the centre was observed (Figure \ref{fig:Jellyfish} D). The ephyrae could regain radial symmetry with up to six arms removed. Even so, development into the medusae stage, where they fully regain swimming functions, could only be possible with at least four remaining arms. Hence radial symmetry also plays a vital role in facilitating further development of the ephyrae \cite{pnas.1502497112}.

The symmetrization process proceeds independently from global factors such as the movement of water, light, or the orientation of the jellyfish in the water column \cite{pnas.1502497112}. The process is also suggested to be independent of wound closure, due to the faster time scale of wound closure, and observing wound closure when symmetrization does not occur. It is important to note that neither cell proliferation nor localized cell death played any important role in the process. Concluding that the process is wholly related to symmetry and musculature, each cycle of contraction and repulsion cause the arms to relax to an increasingly stable state until their morphology is geometrically balanced. Asymmetrical contraction might lead to pivoting of the arms towards the injured sight due to lack of bulk tissue at the site (Figure \ref{fig:Jellyfish} E). Along with the recovery of radial symmetry, this also suggests the existence of a mechanism by which asymmetry is detected \cite{pnas.1502497112}.

Abrams et al. derived a mathematical model for the timescale of symmetry recovery. The model is based on the suggestion that muscular contraction forces and the elastic response of mesoglea involved in the propulsion of the uninjured ephyrae can sufficiently explain the recovery of radial symmetry in injured ephyrae. Using the angular movement of each arm with respect to the geometric centre as a parameter, they apply Hooke's Law to generate a recursive expression for the parameter. The simulation of the mathematical model with parameter values obtained by estimation shows the arms of the ephyrae moving towards the injured site with each contraction and relaxation cycle until symmetry is regained. In other words, symmetry increases with each contraction. Additionally, the model predicts the speed of symmetrization, i.e. how long it takes until the ephyrae regain symmetry, to be dependent on the frequency of muscular contractions. The model suggests that muscular contractions play a dominant role in the mechanism and time required for symmetrization \cite{pnas.1502497112}.

The symmetrization process in jellyfish being a fast self-repair strategy, uses existing physiological machinery to drive the mechanical process. These fast processes prioritize the functional recovery of the organism rather than the recovery of the body parts themselves, which could be possible via regeneration at longer time-scales. Hence, the symmetrization process  could be an energy conserving process, also increasing the animal's survivability. This symmetrization process may inspire biomimetic materials and technologies that require conserving functional geometries without the need for regenerating precise shapes and forms \cite{JELLYFISHBIOIMETIC}. 

\section{Discussion}

\subsection{{\textit{\textit{Form-function Relationships}}}}

In \textit{T. adhaerens}, we have discussed the key role of tissue mechanics in its physiological activities such as reproduction by vegetative fission, and continuous organism-scale shape change. Tissue remodeling mechanisms such as ductile transformations for vegetative reproduction, and brittle deformations for extreme morphological shape changes, are unique adaptations found only in this animal. Hence, these ductile-brittle  tissue transitions have been optimized for their specific flat body plan. Tissue mechanics therefore acts as an important link between the biological form (flat body plan) and function (reproduction and shape change). 

In \textit{H. vulgaris}, we have described how important physiological activities such as feeding and locomotion are intimately determined by tissue mechanics. The tissue remodeling phenomena involved in the mouth opening for feeding and body column bending for movements are unique phenomena found only in this animal. These tissue remodeling phenomena have been optimized for \textit{Hydra}'s specific body plan, possibly suggesting that tissue mechanics links their biological form (body plan) and function (feeding and locomotion).

In \textit{A. aurita}, locomotion in the ephyrae and medusae stage is essential to survival, and we have discussed how there is link between propulsion and maintenance of radial symmetry in these animals. The jellyfish makes use of its radially symmetric musculature to produce power and recovery strokes for propulsion, and these are aided by fast muscle contractions in the arms, and elastic recoil in the viscous mesoglea \cite{JELLYFISHPROPUSION}. The symmetrization process in jellyfish being a fast self-repair strategy thus adopts its existing physiological machinery to drive the mechanical remodeling process. Symmetrization thus encapsulates the priority afforded to shorter time-scale functional recovery over longer time-scale tissue regeneration. Therefore, propulsion and propulsion-linked recovery from amputation, are both tailored for the radial symmetry, tissue flexibility and musculature of the animal.

\subsection{{\textit{\textit{Small-scale Phenomena Lead to Emergent Large-scale Behaviors }}}}

In the ventral epithelium of \textit{T. adhaerens}, micro-scale fractures appear first due to local regions of high tension or shear forces that arise from ciliary-driven traction forces. These micro-fractures subsequently coalesce to form a larger and stable ventral fracture hole. Interestingly, dorsal epithelial fractures also begin with a very small fracture hole that propagates in a different manner and grows in size. Eventually the dorsal hole merges with the ventral hole and an inside edge is created in the animals. Thus, small-scale micro-holes propagate to form larger scale macro-holes in both the ventral and dorsal epithelium.   
\textit{H. vulgaris} mouth opening requires the overcoming of dual sealing forces associated with septate junctions at the cellular scale and contractile myonemes, which act together to enable the large scale phenomena of mouth opening with precise control. Also, localized differential stiffness of tissues in the shoulder and body column give rise to the larger-scale mechanics of the organism's locomotion. Hence, phenomena at the cellular scales determine emergent behavior at the organismal scale. 

\subsection{{\textit{\textit{Rapid Tissue Remodeling versus Regeneration }}}}

This review focuses on rapid (minutes to hours) tissue remodeling phenomena in the three non-bilaterians. However, Cnidarians are also known for their regenerative abilities~\cite{alvarado2006bridging,galliot2019}, and in particular \textit{H. vulgaris} is a popular model system for regeneration. Some Jellyfish species are much more regenerative than the \textit{A. aurita} discussed above, one such example is \textit{Clytia hemisphaerica}.\textit{Clytia hemisphaerica} jellyfish exhibit wound healing and regeneration, and utilize a combination of on tissue re-organization, proliferation of cellular progenitors, and long-range cell recruitment, as shown recently in Ref~\cite{10.7554/eLife.54868}. In these animals, it was found that the rapid tissue remodeling is difficult to separate from the regenerative mechanism in terms of recovery even though the time scales differ~\cite{10.7554/eLife.54868}.

The conflict between the cost-effectiveness of rapid tissue remodeling in relation to regeneration presents tremendous opportunities for further investigation. Additionally, quantifying the amount and type of mechanical forces that can trigger rapid tissue rearrangement and regeneration could shed light on the cost-effectiveness of both processes. Thus, it will be intriguing to look into the small scale, localized molecular mechanisms that translate forces from muscle contraction into tissue re-organization.

\subsection{{\textit{\textit{Tissue Remodeling by Fractures}}}}

\textit{Hydra vulgaris}' mouth opening mechanism by rupture of epithelial tissue is one among the only two known examples of physiological tissue fractures in animals, the other example being the \textit{Trichoplax adhaerens}. However, there are several differences in the tissue fracture phenomena between these two animals. An important difference is the location of tissue fractures, in  \textit{H. vulgaris}, this location is always fixed at the mouth, but in \textit{T. adhaerens}, fractures can arise anywhere in the epithelium -- an emergent phenomenon that depends solely on organismal-scale mechanical forces. Physiological tissue fracture and healing dynamics have not yet been observed in any of the Jellyfish so far, and it could be very interesting to look for them in future work.

Hence, \textit{H. vulgaris} and \textit{T. adhaerens} are powerful model systems for future cell and molecular biology investigations to determine specific components (e.g. proteins/molecules) that enable these unique tissue fractures and their healing dynamics.

\subsection{{\textit{\textit{Tissue Remodeling by Cellular Rearrangements}}}}

Tissue remodeling by local cellular rearrangements is a ubiquitous and well-known mechanism during the development of model animals~\cite{lecuit2013,lecuit2007cell,blankenship2006multicellular,he2014,Mongera2018}. In \textit{T. adhaerens}, we have observed ductile tissue deformations (rapid thinning of tissues) that suggest local cellular rearrangements. In-silico models have indeed revealed the presence of fast cellular rearrangements in the ventral epithelium~\cite{prakash2021motility}, but this has not yet been observed in experiments. 

Unlike \textit{T. adhaerens}, studies of \textit{H. vulgaris} have not reported any fast time-scale cellular rearrangements. Given the dynamic movements of \textit{H. vulgaris}, in future work, it would be interesting to look for localized tissue remodeling events involving cell-cell rearrangements. 

The symmetrization model proposed for recovery in \textit{A. aurita} is based on cellular rearrangements due to forces from contractions that lead to recovery of radial symmetry~\cite{pnas.1502497112}. It was found that the rearrangement of cells is what drives symmetrization, the process being independent of wound healing at the site of injury, ~\cite{pnas.1502497112}.  Despite having different tissue structures and morphologies, other species of jellyfish \cite{ABRAMS20161} and the hydromedusae form \cite{Hydrmedusae} also utilize such a re-organization process in their recovery. 

\subsection{\textit{Tissue Mechanics in Other Non-bilaterian Animals}}\label{subsec9}

We have so far described interesting tissue mechanics phenomena in the phyla of Placozoa and Cnidaria, but non-bilaterians also include two other phyla -- Porifera and Ctenophora. The animals belonging to these two other phyla also have the potential to be good candidates as tissue mechanics models, given their simple body plans and morphologies~\cite{dunn2015hidden}. 

Animals in the phylum Porifera, commonly referred to as sponges, consist of cells in an extracellular matrix, and sometimes also have stiff scaffolding. A recent study dissected their tissues and studied their response to mechanical forces in a rheometer~\cite{Kraus2022}.  It was found that sponge tissues revealed interesting properties, such as anisotropic elasticity, and it was suggested that these properties were linked to the sponge's flow sensitivity~\cite{Kraus2022}. The phylum Ctenophora is characterized by several hundred species of ctenophores, which are soft, gelatinous, freely swimming predatory animals~\cite{pang2008ctenophores}. It has been shown that the ctenophore \textit{Mnemiopsis leidyi} is capable of rapid wound healing when their tissues are cut~\cite{nikki2019,tamm2014cilia}.

There is a huge diversity in Poriferans and Ctenophores, and there are hundreds of different species in both these phyla, with unique adaptations. The few studies on them described above give us a glimpse of the promise of these animals to serve as model systems to investigate tissue remodeling phenomena, particulary tissue rheology and rapid wound healing.

\section{Conclusion}

In this review, we have described interesting tissue mechanics phenomena in two non-bilaterian phyla -- Placozoa and Cnidaria. We have shown how these animals have stretched our perspective and understanding of tissue mechanics. We reveal how animals in these phyla, particularly the Placozoan \textit{Trichoplax adhaerens}, Cnidarians \textit{Hydra vulgaris} and \textit{Aurelia aurita}, utilize rapid tissue remodeling for their physiological functioning~\cite{prakash2021motility,Carter2016,pnas.1502497112} We also illustrate how small-scale cellular reorganization events give rise to phenomena at the larger-scales of tissues. We have shown several examples of how these understudied non-bilaterians present a wealth of opportunities to improve our understanding of rapid tissue remodeling phenomena such as cellular rearrangements, fractures, and wound healing. Hence, these non-bilaterian tissue mechanics model animals complement and contribute to the broader field of biological physics of tissues~\cite{Park2015, Mongera2018, Charras2012,Latorre2018,Xi2019, Charras2022, bull2021b,bull2021c,armon2021,Charras2023,Kim2021,krajnc2020softmatter,Noll2017, Wyatt2016, Bi2015}.

We adopt a comparative approach in this review to allow for a rigorous contribution to the search for general biophysical principles. We hope to have made a strong case for studying these non-model organisms to get to the heart of tissue mechanics and appreciate the extreme examples in order to arrive at a comprehensive framework of tissue mechanics in animals.

\section{Competing interests}
No competing interest is declared.

\section{Author contributions statement}

S.G., G.M., M.R., and V.N.P. discussed ideas, wrote and reviewed the manuscript. V.N.P. conceived the project.

\section{Acknowledgments}
The authors thank several colleagues at the University of Miami for useful discussions and suggestions: Ivan Levkovsky and other students of the Fall 2022 PHY 325/625 (Biological Physics - I) course, members of the Prakash Lab, and William Browne. The authors also thank Matthew Storm Bull, Allen Institute and University of Washington, for valuable feeback and suggestions. V.N.P. thanks Manu Prakash, Stanford University, for introduction to Non-bilaterians and for many stimulating and enriching discussions on tissue mechanics over the years. V.N.P. also thanks the University of Miami for startup funding support.

\bibliographystyle{unsrt}





\end{document}